# Microwave-multiplexed qubit controller using adiabatic superconductor logic

Naoki Takeuchi[1,2,*], Taiki Yamae[3], Taro Yamashita[4], Tsuyoshi Yamamoto[2,5], and Nobuyuki Yoshikawa[3,6]

[1] Global Research and Development Center for Business by Quantum-AI Technology, National Institute of Advanced Industrial Science and Technology (AIST), Tsukuba, Ibaraki 305-8568, Japan
[2] NEC-AIST Quantum Technology Cooperative Research Laboratory, National Institute of Advanced Industrial Science and Technology (AIST), Tsukuba, Ibaraki 305-8568, Japan
[3] Institute of Advanced Sciences, Yokohama National University, Yokohama, Kanagawa 240-8501, Japan
[4] Department of Applied Physics, Tohoku University, Sendai, Miyagi 980-8579, Japan
[5] Secure System Platform Research Laboratories, NEC Corporation, Kawasaki, Kanagawa 211-0011, Japan
[6] Department of Electrical and Computer Engineering, Yokohama National University, Yokohama, Kanagawa 240-8501, Japan

[*] E-mail: n-takeuchi@aist.go.jp

Cryogenic qubit controllers (QCs) are the key to build large-scale superconducting quantum processors. However, developing scalable QCs is challenging because the cooling power of a dilution refrigerator is too small (~10 µW at ~10 mK) to operate conventional logic families, such as complementary metal-oxide-semiconductor logic and superconducting single-flux-quantum logic, near qubits. Here we report on a scalable QC using an ultra-low-power superconductor logic family, namely adiabatic quantum-flux-parametron (AQFP) logic. The AQFP-based QC, referred to as the AQFP-multiplexed QC (AQFP-mux QC), produces multi-tone microwave signals for qubit control with an extremely small power dissipation of 81.8 pW per qubit. Furthermore, the AQFP-mux QC adopts microwave multiplexing to reduce the number of coaxial cables for operating the entire system. As a proof of concept, we demonstrate an AQFP-mux QC chip that produces microwave signals at two output ports through microwave multiplexing and demultiplexing. Experimental results show an output power of approximately −80 dBm and on/off ratio of ~40 dB at each output port. Basic mixing operation is also demonstrated by observing sideband signals.

## INTRODUCTION

A significant engineering effort is required to develop practical, fault-tolerant quantum processors (QPs). Quantum computing has the potential to surpass classical computing in some applications[1–4]; however, quantum error correction requires numerous physical qubits[5–8]. This is a great engineering challenge for superconducting QPs because superconducting qubits are cooled to ~10 mK inside a dilution refrigerator to suppress thermal noise and should be controlled in a hardware-efficient way[9]. Currently, superconducting QPs[10,11] (including hundreds of qubits at the most[12]) use the brute-force scheme to control qubits, in which microwave pulses generated by room-temperature electronics are applied to each qubit via coaxial cables between 300-K and 10-mK stages. This control scheme is not scalable because the number of available coaxial cables is limited by the cooling power and physical space of the dilution refrigerator[13].

One of the most promising control schemes is the implementation of cryogenic qubit controllers (QCs) in dilution refrigerators. Recently, two types of QCs have been developed extensively, cryogenic complementary metal-oxide-semiconductor (cryo-CMOS) logic-based[14-16] and superconducting single-flux-quantum (SFQ) logic-based[17–19] QCs. Although both QCs successfully controlled superconducting qubits, they need to be placed at relatively warm stages (i.e., the 3-K stage); therefore, they still have scalability limitations associated with coaxial cables between the 3-K and 10-mK stages. This is because the cooling power of the 10-mK stage is too small (~10 µW) to operate cryo-CMOS and SFQ circuits, whereas a very simple SFQ QC can operate at the 10-mK stage[17]. Typically, the power dissipation of cryo-CMOS and SFQ QCs is 1–100 mW per qubit[16] and 1–100 µW per qubit[18,19], respectively. Therefore, to build large-scale superconducting QPs, it is crucial to develop ultra-low-power QCs that can operate together with qubits at the 10-mK stage.

In this study, we propose a scalable QC using adiabatic superconductor logic, namely adiabatic quantum-flux-parametron (AQFP) logic[20,21]. AQFP circuits are based on the quantum flux parametron[22,23] and can operate with an extremely small power dissipation of ~7 pW per Josephson junction[24] owing to adiabatic switching[25–27]. For instance, a very-large-scale AQFP circuit having $10^6$ Josephson junctions can operate with a power dissipation of only ~7 μW, which is still less than the cooling power of the 10-mK stage. Furthermore, inspired by microwave-multiplexed superconducting sensor arrays[28,29], the proposed QC adopts AQFP-based microwave multiplexing to reduce the number of coaxial cables for driving the QC. These features ensure very high scalability in terms of both power dissipation and cable count, indicating the possibility of implementation with qubits at the 10-mK stage. The proposed QC is termed the AQFP-multiplexed QC (AQFP-mux QC).

In the following sections, we describe the details of the AQFP-mux QC using numerical simulations and demonstrate an AQFP-mux QC chip at 4.2 K as a proof of concept. Specifically, we show that the AQFP-mux QC can produce multi-tone microwave signals for controlling qubits with an ultra-low power dissipation of 81.8 pW per qubit and can be driven by using a single coaxial cable, independently of the qubit count. For clear terminology, we distinguish AQFP logic from SFQ logic, since typical SFQ logic families (such as rapid SFQ logic[30]) use SFQ pulses to encode data, whereas AQFP logic uses the polarity of current for data encoding.

## RESULTS
### AQFP-multiplexed qubit controller

Fig. 1a illustrates a conceptual diagram of the AQFP-mux QC, where both the AQFP-mux QC and qubit chips are implemented at the 10-mK stage inside a dilution refrigerator. A multi-tone microwave generated by room-temperature electronics is applied to the AQFP-mux QC chip via a single coaxial cable connecting the 300-K and 10-mK stages. The AQFP-mux QC demultiplexes the microwave tones and applies microwave pulses with different frequencies to the qubits, where the qubit frequencies should match the microwave tones. The number of coaxial cables in the entire system does not increase with the qubit count owing to microwave multiplexing, which is a clear advantage over previously proposed qubit controllers[14–19]. The AQFP-mux QC comprises both analog and digital parts. The analog component demultiplexes the input microwave tones, mixes each microwave tone with a baseband signal to produce microwave pulse trains, and switches each microwave pulse on and off. The digital component sends digital signals to the analog component to control the switching of each microwave pulse in accordance with the given quantum algorithm. Because the main component of the AQFP-mux QC is the analog part, we will describe only the analog part in this study and discuss the digital part in future work.

The core of the AQFP-mux QC is the AQFP mixer shown in Fig. 1b, which produces a microwave pulse train by mixing the local oscillator (LO) and baseband (BB) signals and switches each microwave pulse on/off following the digital inputs. The AQFP mixer comprises paired AQFPs (A and B) coupled to two excitation currents, $I_{lo}$ and $I_{bb}$. $I_{lo}$ and $I_{bb}$ apply LO and BB magnetic fluxes with an amplitude of $0.5\Phi_0$ each, to each AQFP, where $\Phi_0$ is the flux quantum. Paired AQFPs are excited when both $I_{lo}$ and $I_{bb}$ are at high levels, consequently generating mixed signals of $I_{lo}$ and $I_{bb}$ as output currents ($I_{outa}$ and $I_{outb}$). $I_{outa}$ and $I_{outb}$ apply magnetic fluxes ($\Phi_a$ and $\Phi_b$) to the load line, the time derivative of which generates electromotive force and appears as the output microwave $V_{out}$ through a bandpass filter (BPF). The switching of $V_{out}$ is controlled by digital input currents $I_{in}$ and $I_{fix}$ (i.e., $V_{out}$ is switched on/off by the logical states of the paired AQFPs). $V_{out}$ is turned on when $I_{in}$ and $I_{fix}$ have the same logical values, because $\Phi_a$ and $\Phi_b$ have the same polarity. In contrast, $V_{out}$ is turned off when $I_{in}$ and $I_{fix}$ have different logical values, because $\Phi_a$ and $\Phi_b$ cancel out each other. Hereafter, $I_{fix}$ is fixed to logic 1; therefore, the switching of $V_{out}$ is controlled by $I_{in}$.

Fig. 1c shows the numerical simulation of the AQFP mixer by JoSIM[31], where $I_{lo}$ is a 5-GHz sinusoidal current, $I_{bb}$ is a triangular current as an example of BB signals, and the dashed lines represent zero for each waveform. The figure clearly shows that the AQFP mixer generates a 5-GHz microwave pulse ($V_{out}$) by mixing $I_{lo}$ and $I_{bb}$, and switches on/off $V_{out}$ by $I_{in}$ (the polarities of $I_{in}$ and $I_{fix}$ represent their logical values). The envelope of $V_{out}$ is not identical to the shape of $I_{bb}$ because of the nonlinear relationship between $I_{bb}$ and the output currents ($I_{outa}$ and $I_{outb}$), that is, the AQFP mixer operates as a nonlinear mixer. Precise pulse shaping for $V_{out}$ is possible by considering the nonlinear relationship, as shown later.

Fig. 1d shows a detailed diagram of the AQFP-mux QC. Multiple AQFP mixers are excited by a single LO current ($I_{lo}$) including multiple microwave tones, using a superconducting resonator array as a microwave demultiplexer. Assuming that $I_{lo}$ includes three different microwave tones ($f_1$, $f_2$, and $f_3$) that correspond to the qubit frequencies, AQFP mixers 1, 2, and 3 are coupled to $I_{lo}$ through resonators with resonance frequencies of $f_1$, $f_2$, and $f_3$, respectively. Furthermore, all AQFP mixers are coupled to the common BB current $I_{bb}$. Consequently, each AQFP mixer generates microwave pulses with a different frequency by mixing $I_{bb}$ and the LO signal in the resonator coupled to the mixer. For instance, AQFP mixer 1 generates microwave pulses $V_{out1}$ with a frequency of $f_1$ by mixing $I_{bb}$ and the LO current in the $f_1$ resonator ($I_{r1}$). Importantly, the number of control lines ($I_{lo}$ and $I_{bb}$) does not increase with the qubit count because all AQFP mixers share $I_{lo}$ and $I_{bb}$ lines. Note that the level of $I_{lo}$ for each tone is determined such that magnetic flux with an amplitude of $0.5\Phi_0$ is applied to each AQFP in the mixers (see the Methods section for more details).

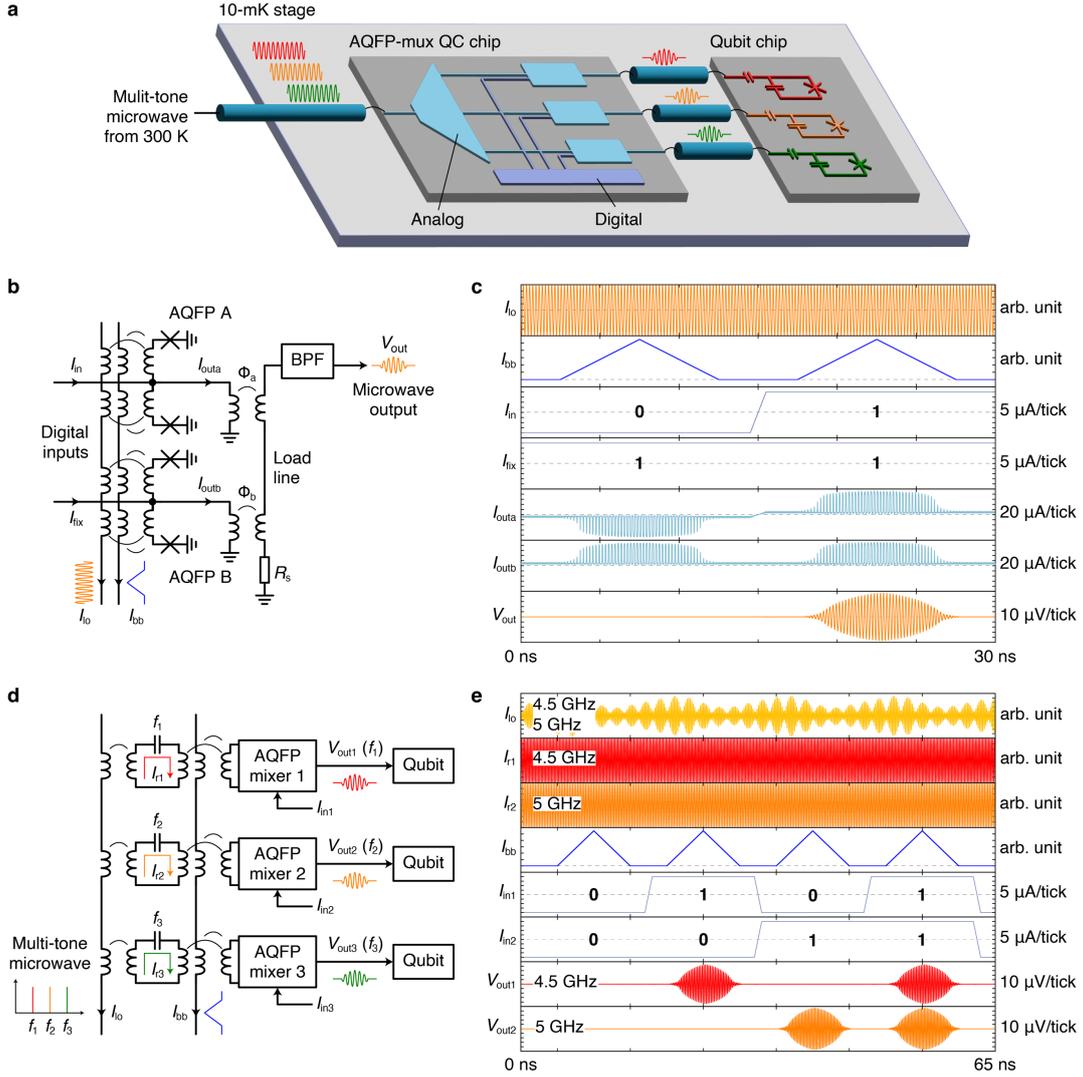

**Fig. 1 | AQFP-multiplexed qubit controller (AQFP-mux QC). a.** Conceptual diagram. The AQFP-mux QC chip produces microwave pulses for controlling qubits from a single multi-tone microwave input at the 10-mK stage, thereby solving the scalability limitation due to coaxial cables between 300-K and 10-mK stages. **b.** AQFP mixer. The output voltage $V_{out}$ is generated by mixing the local-oscillator (LO) current $I_{lo}$ and baseband current $I_{bb}$ using the nonlinearity of AQFPs A and B. **c.** Numerical simulation of the AQFP mixer. A microwave pulse is generated at $V_{out}$ by mixing $I_{lo}$ (5-GHz sinusoidal current) and $I_{bb}$ (triangular current as an example), and $V_{out}$ is switched on/off by the digital input currents $I_{in}$ and $I_{fix}$. **d.** Detailed diagram of the AQFP-mux QC. Multiple AQFP mixers are driven by a single multi-tone LO current $I_{lo}$ using a superconducting resonator array operating as a microwave demultiplexer. The number of control lines ($I_{lo}$ and $I_{bb}$) does not increase with the qubit count thanks to microwave multiplexing. **e.** Numerical simulation of the AQFP-mux QC, where $I_{lo}$ includes two microwave tones (4.5 GHz and 5 GHz). 4.5-GHz and 5-GHz microwave pulses are generated at different output ports ($V_{out1}$ and $V_{out2}$) and individually switched on/off by the input currents $I_{in1}$ and $I_{in2}$, respectively.

Fig. 1e shows the numerical simulation of the AQFP-mux QC by JoSIM, where $I_{lo}$ includes two microwave tones ($f_1$ = 4.5 GHz and $f_2$ = 5 GHz). $I_{in1}$ and $I_{in2}$ are the input currents applied to AQFP mixers 1 and 2, respectively, to switch the outputs on and off. The figure shows that the two AQFP mixers generate microwave pulses ($V_{out1}$ and $V_{out2}$) with different frequencies ($f_1$ and $f_2$), and that each output is individually turned on/off by $I_{in1}$ or $I_{in2}$. This result indicates that the AQFP-mux QC can apply microwave pulses to arbitrary qubits in accordance with the given quantum algorithm.

## Performance estimation

We estimated the performance (output power and power dissipation) of the AQFP mixer through numerical simulations using JSIM_n[32]. The performance of the AQFP mixer changes with the load line design (Fig. 1b). In this study, the impedance of the load line was designed to be a small value of 2 Ω to extract relatively large output power from the mixer; accordingly, the matching resistor $R_s$ and input impedance of the BPF were designed to be 2 Ω. The BPF was also designed to convert impedance from 2 Ω (input) to 50 Ω (output) to drive general 50-Ω loads. Details of the BPF design can be found in the Methods section. In both the numerical simulation and experiments, the AQFP mixer was evaluated with a 50-Ω load at the output port.

The output power (when $I_{bb}$ was at the maximum) was estimated to be 21.9 pW (= −76.6 dBm) for 5-GHz operation, which is assumed to be large enough for typical qubit control. If necessary, the output power can be further increased by adding more AQFPs to the load line of the AQFP mixer. Two types of power dissipation were estimated for the 5-GHz operation, namely standby power $P_{sby}$ (when $I_{bb}$ was off) and maximum power $P_{max}$ (when $I_{bb}$ was at the maximum). $P_{sby}$ and $P_{max}$ were estimated to be 2.82 pW (2.29 pW for AQFPs and 0.533 pW for $R_s$) and 81.8 pW (22.3 pW for AQFPs and 59.5 pW for $R_s$), respectively. Hence, the power dissipation of the AQFP-mux QC is 81.8 pW per qubit at the most, excluding the digital processing circuits. This power dissipation is orders of magnitude lower than cryo-CMOS and SFQ QCs. Note that the power dissipation of the AQFP-mux QC was estimated using circuit parameters at 4.2 K since the circuit parameters do not vary much between 4.2 K and 10 mK. For instance, the change in the critical current of an Nb-based Josephson junction (critical temperature: ~9 K) is estimated to be less than 10% between 4.2 K and 10 mK[33,34]. Also, this parameter change can be compensated by slightly changing the layout design of each junction when customizing the AQFP-mux QC for 10-mK operation.

In addition, we discuss the frequency efficiency of the microwave multiplexing. Assuming the use of Nb resonators[35] with a quality factor $Q$ of ~$10^4$ and a resonance frequency $f_0$ of ~5 GHz, the minimum frequency spacing between adjacent microwave tones is approximately given by $f_0/Q$ = 500 kHz. Therefore, a single LO line can transmit ~4,000 microwave tones, assuming a bandwidth of 2 GHz. This is much more frequency-efficient than the frequency multiplexing used in the cryo-CMOS QC[15], which can multiplex 32 signals over a control line. This is because continuous microwaves are multiplexed in the AQFP-mux QC, whereas microwave pulses are multiplexed in the cryo-CMOS QC (i.e., wide frequency spacing is required).

We roughly compare the scalability of the AQFP-mux QC with that of other QCs from the viewpoints of power dissipation and the number of coaxial cables. According to the literature[15], the power dissipation of a cryo-CMOS QC is 12 mW per qubit. The cryo-CMOS QC is placed at the 3-K stage, and each signal line can multiplex 32 qubits by frequency multiplexing; thus, the number of the coaxial cables between 10-mK and higher-temperature (i.e., 3-K) stages is $N_{cable}$ ~ $N_{qubit}$/32 ($N_{qubit}$: number of qubits), where the cables for two-qubit gates and readout are omitted for simplicity. The power dissipation of SFQ QCs is 1.6 μW per qubit and 51.7 μW per qubit for the pulse-based control[18] and microwave-based control[19], respectively. The SFQ QCs are assumed to be placed at 3-K stages and currently do not implement frequency-multiplexing, i.e., $N_{cable}$ ~ $N_{qubit}$. The power dissipation of the AQFP-mux QC is 81.8 pW per qubit, and each LO line can multiplex ~4,000 qubits owing to microwave multiplexing, i.e., $N_{cable}$ ~ $N_{qubit}$/4000. The above comparison indicates that the AQFP-mux QC has the potential to control the qubits in a large-scale QP with much less power dissipation and cable count than other QCs. It should also be noted that, at this moment, the cryo-CMOS QC has much richer circuit functions than the SFQ and AQFP-mux QCs; thus, more systematic, fair comparison will be required in future work.

**Experimental demonstration**

As a proof of concept, we fabricated an AQFP-mux QC chip and demonstrated its operation at 4.2 K in liquid helium using a wideband cryoprobe[36]. As mentioned above, the changes in the circuit parameters between 4.2 K and 10 mK are small and do not make a difference on the operating principle of the AQFP-mux QC; thus, we conducted proof-of-concept experiments at 4.2 K. Fig. 2a shows a micrograph of the AQFP-mux QC chip, which multiplexes two AQFP mixers (1 and 2) using two resonators with resonance frequencies of $f_1$ and $f_2$. Both the mixers were excited by a single $I_{lo}$ including two microwave tones ($f_1$ and $f_2$) and a common $I_{bb}$. The outputs of the mixers ($V_{out1}$ and $V_{out2}$) were controlled by the input currents ($I_{in1}$ and $I_{in2}$) and were observed using a signal analyzer. We analyzed the insertion loss of the LO line using a vector network analyzer to determine the resonance frequencies and found $f_1$ = 4.3392 GHz and $f_2$ = 4.8171 GHz.

First, we demonstrated the switching operation of the AQFP-mux QC. A DC current inducing $0.5\Phi_0$ on each AQFP in the mixers was applied to $I_{bb}$, and $V_{out1}$ and $V_{out2}$ were observed for different sets of the logical values of $I_{in1}$ and $I_{in2}$ ($a_1$ and $a_2$). Fig. 2b shows the measurement results of $V_{out1}$ at $f_1$ and $V_{out2}$ at $f_2$ for $(a_1, a_2) \in \{(0, 0), (0, 1), (1, 0), (1, 1)\}$. This figure indicates that $V_{out1}$ and $V_{out2}$ can be individually controlled by $I_{in1}$ and $I_{in2}$, respectively, as shown in the waveforms in Fig. 1e. For instance, $V_{out1}$ is switched on for (1, 0) and (1, 1), and off for (0, 0) and (0, 1). The output power of $V_{out1}$ and $V_{out2}$ when switched on is −82.1 dBm and −81.8 dBm, respectively. These power values may be smaller than the estimated value (−76.6 dBm) because of cryoprobe losses and parameter mismatches between the design values and fabricated chip. The on/off ratios of $V_{out1}$ and $V_{out2}$ are 42.7 dB and 39.0 dB, respectively, which are comparable with those of previously reported cryogenic microwave switches[37–40]. We also evaluated the leakage power between the output channels, that is, $V_{out1}$ at $f_2$ and $V_{out2}$ at $f_1$, as shown in Fig. 2c. The leakage power to $V_{out1}$ and $V_{out2}$ is at the most −117.1 dBm and −121.1 dBm, respectively. The on/off ratios and leakage power may have been determined by various factors, such as crosstalk between $I_{lo}$ and the output channels inside the cryoprobe, parameter imbalance in the AQFPs, and interaction between the AQFP mixers. Further studies are required to improve the on/off ratios and reduce the leakage power.

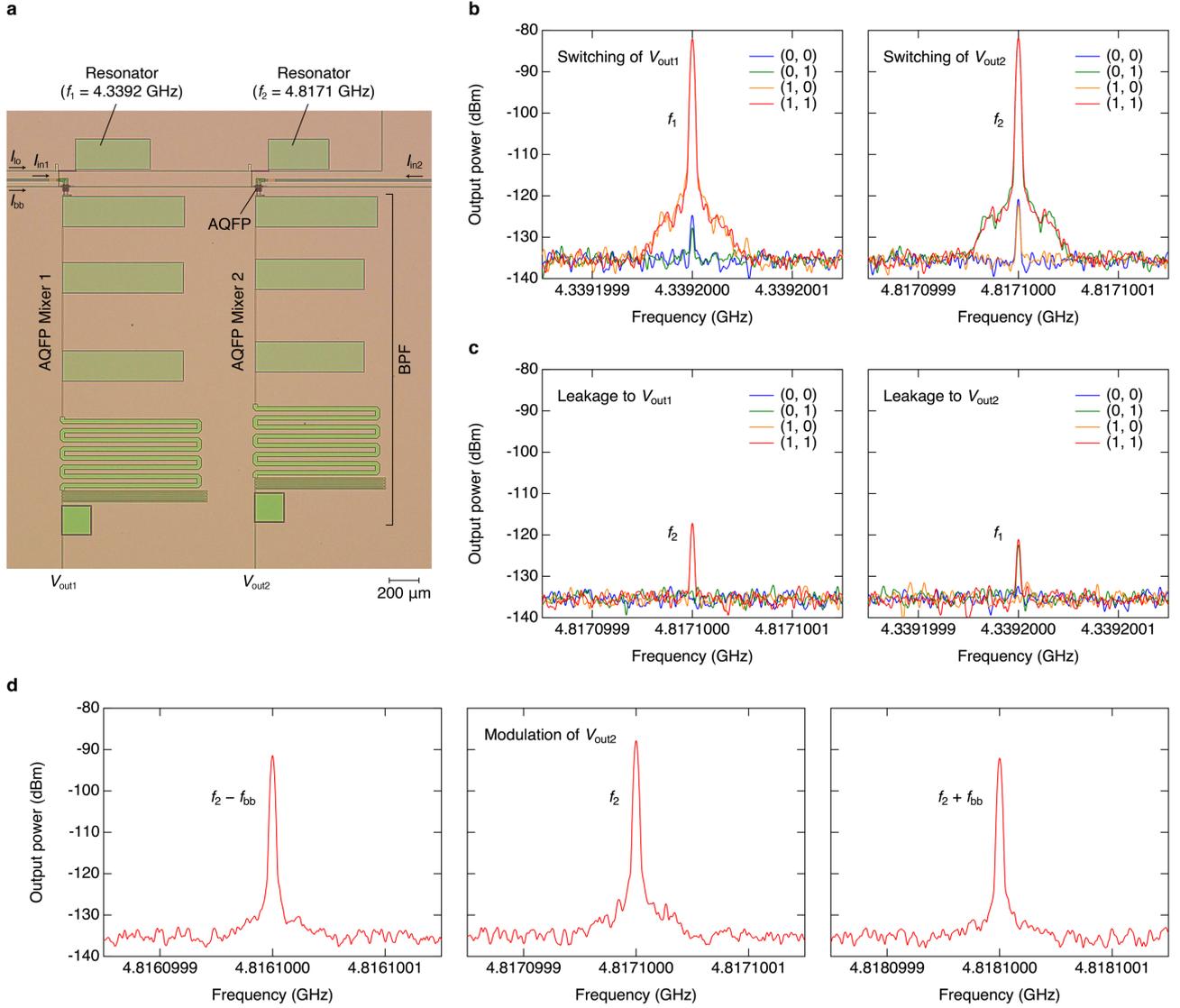

**Fig. 2 | Experimental demonstration. a.** Micrograph of an AQFP-mux QC chip, multiplexing two AQFP mixers (1 and 2) via two resonators with resonance frequencies of $f_1 = 4.3392$ GHz and $f_2 = 4.8171$ GHz. The mixers were excited by a single LO current $I_{lo}$ including $f_1$ and $f_2$ tones. The output voltages from the mixers ($V_{out1}$ and $V_{out2}$) were observed by a signal analyzer. **b.** Switching operation for $V_{out1}$ at $f_1$ and $V_{out2}$ at $f_2$, with the baseband current $I_{bb}$ fixed to induce $0.5\Phi_0$ on each AQFP. $V_{out1}$ and $V_{out2}$ are individually switched on/off by $I_{in1}$ and $I_{in2}$, respectively, the logical values of which are represented by $(a_1, a_2)$. The output power of $V_{out1}$ and $V_{out2}$ when switched on is −82.1 dBm and −81.8 dBm, respectively. The on/off ratios of $V_{out1}$ and $V_{out2}$ are 42.7 dB and 39.0 dB, respectively. **c.** Leakage power between output channels, i.e., $V_{out1}$ at $f_2$ and $V_{out2}$ at $f_1$. The leakage power to $V_{out1}$ and $V_{out2}$ is at the most, −117.1 dBm and −121.1 dBm, respectively. **d.** Mixing operation for $V_{out2}$, with a 1-MHz square current applied to $I_{bb}$. Peak power appears at the LO and sideband frequencies (i.e., $f_2$ and $f_2 \pm 1$ MHz), demonstrating modulation by AQFP mixer 2.

We then demonstrated the basic mixing operation of the AQFP-mux QC by observing sideband signals generated from LO and baseband signals. A square pulse with a repetition frequency $f_{bb}$ of 1 MHz was applied to $I_{bb}$ (the low and high levels induced $0\Phi_0$ and $0.5\Phi_0$, respectively, on each AQFP), and $V_{out1}$ and $V_{out2}$ were observed at the LO and sideband frequencies for $(a_1, a_2) = (1, 1)$. We randomly selected a repetition frequency of 1 MHz since any frequencies below the LO frequency can be used for sideband observation. Fig. 2d shows $V_{out2}$ at $f_2$ and $f_2 \pm f_{bb}$, which demonstrates that the $I_{bb}$ signal is upconverted by $f_2$, i.e., AQFP mixer 2 modulates $V_{out2}$ by mixing $I_{bb}$ and the $f_2$ tone in $I_{lo}$. We observed similar modulation characteristics for $V_{out1}$. It should be noted that the duration time of qubit-control pulses is tens of nanoseconds[41], so we plan to perform much faster modulation in future experiments to enhance the feasibility of AQFP-based qubit control.

## Power calibration

In addition to the individual switching of each output, QCs must implement the individual power calibration of each output because the output power of each mixer and the coupling strength between each pair of output channels and qubits can vary from device to device and chip to chip. The AQFP-mux QC can perform individual power calibrations by adjusting each microwave tone in $I_{lo}$. Fig. 3a shows the simulated waveforms for the AQFP-mux QC when the power of the $f_2$ tone was reduced by 3 dB compared with the normal excitation condition (Fig. 1e). A comparison of Figs. 1e and 3a shows that the amplitude and duration of the microwave pulses at $V_{out2}$ decrease with decreasing $f_2$ tone power in $I_{lo}$ (the duration decreases because a larger $I_{bb}$ value is required to excite the AQFPs), whereas those at $V_{out1}$ do not change, that is, the power of $V_{out2}$ can be individually adjusted without changing that of $V_{out1}$.

We experimentally demonstrated individual power calibration using an AQFP-mux QC chip. Fig. 3b shows the power of $V_{out1}$ and $V_{out2}$ as a function of the $f_2$ tone power in $I_{lo}$, where an $f_2$ tone of −56 dBm is the normal excitation condition and a DC current inducing $0.5\Phi_0$ on each AQFP was applied to $I_{bb}$. The figure shows that the power of $V_{out2}$ can be individually adjusted by the $f_2$ tone power over a wide range while keeping the power of $V_{out1}$ almost constant. We observed similar individual power calibrations for $V_{out1}$.

## Pulse shaping

It is important to apply microwave pulses with an appropriate envelope to each qubit to avoid unintentional qubit excitation[41], such as leakage to a higher energy level. As mentioned above, the AQFP mixer operates as a nonlinear mixer; therefore, special care should be taken for $I_{bb}$ to achieve $V_{out}$ with the desired envelope. Precise pulse shaping is possible by designing the evolution of $I_{bb}$ considering the nonlinear relationship between $V_{out}$ and $I_{bb}$. Fig. 4a shows the simulation results for the amplitude of $V_{out}$ as a function of the magnetic flux $\Phi_{bb}$ applied to each AQFP in the mixer by $I_{bb}$. The amplitude of $V_{out}$ is determined by $\Phi_{bb}$ one-to-one, so the time evolution of $I_{bb}$ for achieving a desired envelop is uniquely determined from Fig. 4a. Figs. 4b and 4c show the simulated waveforms that achieved $V_{out}$ with triangular and Gaussian envelopes, respectively, as examples of pulse shaping. The envelope of $V_{out}$ was first determined, and the evolution of $I_{bb}$ was designed by converting $V_{out}$ at each time step to the corresponding $I_{bb}$ value, based on Fig. 4a. These results suggest that the AQFP mixers in the AQFP-mux QC can produce microwave pulses with appropriate envelopes by carefully designing the evolution of $I_{bb}$.

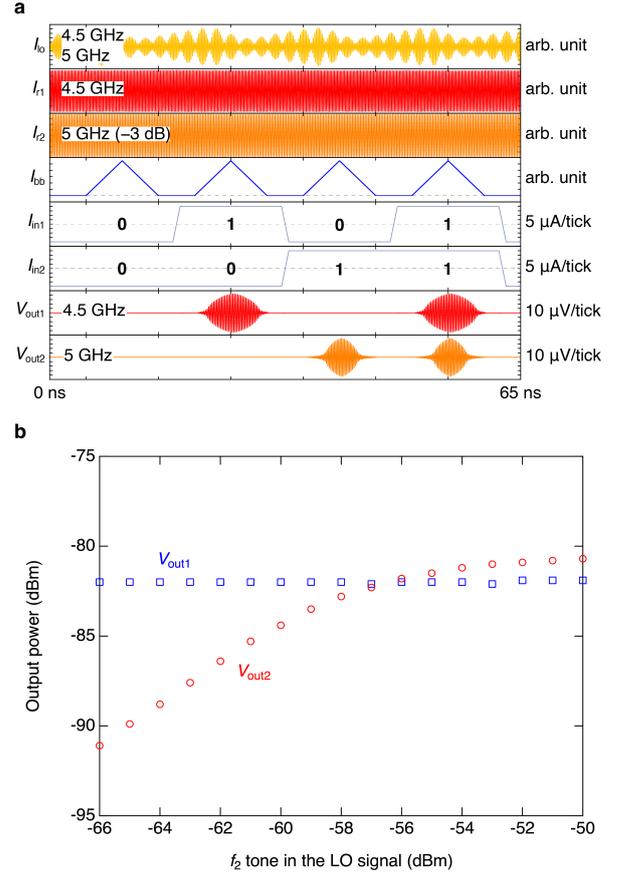

**Fig. 3 | Power calibration. a.** Numerical simulation of the AQFP-mux QC when the power of the $f_2$ tone (5 GHz) in the LO current $I_{lo}$ is reduced by 3 dB, compared to the numerical simulation shown in Fig. 1e. The amplitude and duration of the microwave pulses at $V_{out2}$ decrease with decreasing $f_2$ tone power. **b.** Measurement results of each output power vs the $f_2$ tone power in $I_{lo}$. The power of $V_{out2}$ can be individually adjusted by the $f_2$ tone power, without changing that of $V_{out1}$.

## DISCUSSION

We discuss the challenges of integration with qubits. In the AQFP-mux QC, the qubit frequencies are assumed to be equal to the resonance frequencies of the resonator array operating as a microwave demultiplexer, even with process variation. Therefore, a frequency-matching mechanism must be incorporated. It is not possible to use superconducting quantum interference device (SQUID)-based tunable resonators[42] for frequency matching because a relatively large current must flow through a resonator to drive an AQFP mixer. Therefore, integration with frequency-tunable qubits such as flux qubits[43] and split transmons[44] is preferable. To calibrate each output power individually, it is also assumed that all qubit frequencies are different from each other. This constraint must be considered when designing the qubit frequencies.

Furthermore, circuit design challenges should be considered. A QC must control the phases of the produced microwave pulses to form a universal quantum gate set[41]. For instance, the "Clifford + $T$" set[45] includes a $T$ gate, which rotates the qubit state by 45 degrees about the $Z$-axis; in this case, a QC needs to control microwave phases by 45 degrees, assuming the use of virtual $Z$ gates[46]. Currently, the AQFP mixers in the AQFP-mux QC can control the microwave phases by only 180 degrees [in Fig. 1c, the phase of $V_{out}$ changes by 180 degrees by inverting the polarities of $I_{in}$ and $I_{fix}$]. Thus, the analog part of the AQFP-mux QC requires improvement to achieve a better phase resolution. Moreover, the digital part should be carefully designed so that the entire architecture exhibits high scalability. One of the most straightforward designs for the digital part is an address decoder[37] that selects which AQFP mixers to turn on following externally applied digital signals. Regardless of the architecture adopted, it is crucial to densely combine analog and digital parts using high-density circuit technology, such as a multi-junction-layer process[47] and multi-chip module[48]. It will be also necessary to investigate how to integrate the AQFP-mux QC with memory circuits that store pulse sequence information and qubit readout circuits.

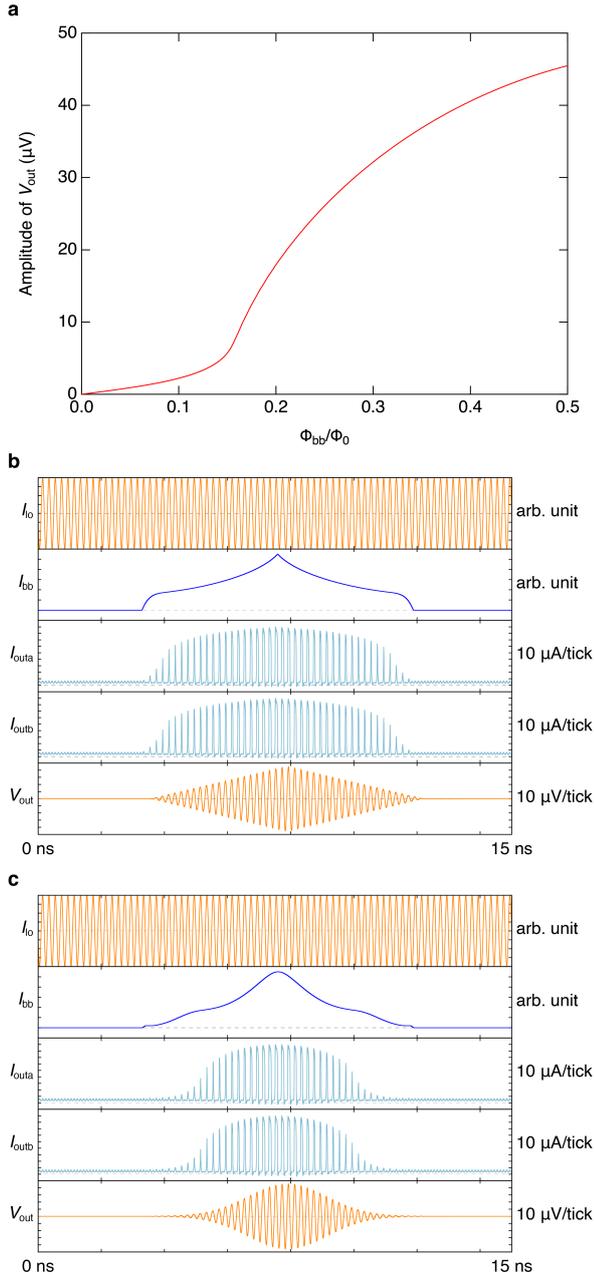

**Fig. 4 | Pulse shaping. a.** Numerical simulation for $V_{out}$ vs magnetic flux applied to each AQFP ($\Phi_{bb}$) by the baseband current $I_{bb}$. **b.** Triangular microwave pulse. **c.** Gaussian microwave pulse. Microwave pulses with desired envelopes can be achieved by designing the evolution of $I_{bb}$ based on the nonlinear relation shown in Fig. 4a.

In conclusion, we proposed the AQFP-mux QC as a scalable QC for large-scale superconducting QPs. The AQFP mixers in the AQFP-mux QC produce microwave pulse trains for controlling qubits by exploiting the nonlinearity of AQFPs, and switch each microwave pulse on/off by controlling the output polarities of the AQFPs. The AQFP-mux QC operates with an ultra-low power dissipation of 81.8 pW per qubit at the most, except for the digital processing circuits, and can be driven by a frequency-multiplexed microwave using a single coaxial cable. These features suggest the high scalability of the AQFP-mux QC in terms of both power dissipation and cable count. As a proof-of-concept, we demonstrated the operation of an AQFP-mux QC chip at 4.2 K, which multiplexed two AQFP mixers using a two-tone microwave. The experimental results indicated an output power of approximately −80 dBm and an on/off ratio of approximately 40 dB at each output port. The basic mixing operation was also demonstrated by observing LO and sideband signals. These results indicate the feasibility of energy-efficient, scalable qubit control using AQFP logic. Our next step is to demonstrate the AQFP-mux QC chip at 10 mK (where fast modulation toward qubit control should be performed) as a preliminary test for integration with qubits.

## METHODS

**AQFP design.** The AQFPs in the AQFP mixer were designed using 50-μA Josephson junctions and inductances based on previously developed AQFP cell libraries. The physical layout was designed by estimating each inductance using InductEx[49].

**BPF design.** The BPF in the AQFP mixer comprises a low-pass filter (LPF), impedance transformer, and DC blocking capacitor (3 pF) in series. The LPF is a five-pole Chebyshev LPF with an impedance of 2 Ω, passband ripple of 0.2 dB, and cutoff frequency of $1.3f_{op}$, where $f_{op}$ is the operating frequency. The impedance transformer is a two-section Chebyshev quarter-wave transformer[50], which converts an input impedance of 2 Ω to an output impedance of 50 Ω with a bandwidth of $0.4f_{op}$. In the AQFP-mux QC chip, the BPFs in AQFP mixers 1 and 2 were designed for $f_{op}$ = 4.5 GHz and 5 GHz, respectively.

**Resonator design.** The resonators used for microwave demultiplexing are lumped-element $LC$ resonators. Each resonator was designed using the resonator model shown in Supplementary Fig. 1a, where $R$ represents the equivalent resistance of the source and load impedances for the LO line (i.e., $R = 50\ \Omega + 50\ \Omega = 100\ \Omega$), and $L_r$ and $C_r$ are the resonator inductance and capacitance, respectively. The resonator is magnetically coupled to the LO line through a transformer composed of $L_1$ and $L_2$ with mutual inductance $M = k(L_1 L_2)^{0.5}$. The resistance of the resonator was assumed negligible, that is, the loaded quality factor was equal to the external quality factor. Supplementary Fig. 1b is the equivalent circuit after converting the LO line using $M$ as an impedance inverter, where $R' = (\omega M)^2 R/[R^2 + (\omega L_1)^2]$, $L_1' = -(\omega M)^2 L_1/[R^2 + (\omega L_1)^2]$, and $\omega$ is an angular frequency. From this circuit diagram, the resonance angular frequency is given by $\omega_0 = 1/(L_{tot} C_r)^{0.5}$, where $L_{tot} = L_r + L_1' + L_2$, and the quality factor is given by $Q = \omega_0 L_{tot}/R' = \{[R^2 + (\omega_0 L_1)^2] L_r + [R^2 + (1 - k^2)(\omega_0 L_1)^2] L_2\}/\omega_0 M^2 R$. Because of the power equivalence between the two circuit diagrams ($R I_{lo}^2/2 = R' I_r^2/2$), the current ratio is given by $I_r/I_{lo} = [R^2 + (\omega_0 L_1)^2]^{0.5}/\omega_0 M$, where $I_{lo}$ is the amplitude of the current through the LO line and $I_r$ is that through the resonator. Assuming $R^2 \gg (\omega_0 L_1)^2$, $\omega_0 \approx 1/[(L_r + L_2) C_r]^{0.5}$, $Q \approx (L_r + L_2) R/\omega_0 M^2$, and $I_r/I_{lo} \approx R/\omega_0 M$. Resonators 1 and 2 in the AQFP-mux QC chip were designed for $\omega_0/2\pi = 4.5$ GHz and $\omega_0/2\pi = 5.0$ GHz, respectively, with a $Q$ of ~200. The power of each microwave tone was determined using the current ratios ($I_r/I_{lo} = 88$ and $I_r/I_{lo} = 79$ for resonators 1 and 2, respectively).

**Power calculation.** The entire power dissipation $P_{tot}$ of the AQFP mixer plus output load (50 Ω) was calculated by integrating the current through and voltage across the LO line over time[51], and the output power $P_{out}$ was calculated by integrating the current through and voltage across the output load over time. The power dissipation of the AQFP mixer was then calculated by $P_{aqfp} = P_{tot} - P_{out}$.

**Circuit fabrication.** The AQFP-mux QC chip was fabricated using a 10-kA/cm$^2$ four-Nb-layer process provided by AIST, namely the high-speed standard process[52].

**Measurement.** The LO signal ($I_{lo}$), which included two microwave tones ($f_1$ and $f_2$), was generated by combining microwaves from a two-channel signal generator (Anritsu, MG3710A) using a power combiner. The BB signal ($I_{bb}$) was generated using an arbitrary waveform generator (Keysight, M3202A). The digital inputs ($I_{in1}$ and $I_{in2}$) were generated using a multichannel source measurement unit (Keysight, M9614A). The outputs of the AQFP-mux QC chip ($V_{out1}$ and $V_{out2}$) were directly observed without amplification using a signal analyzer (Anritsu, MS2830A).

## Data availability

The data that support the findings of this study are available from the corresponding author upon reasonable request.

## Acknowledgements
This study was supported by the JST FOREST (Grant No. JPMJFR212B) and JSPS KAKENHI (Grants No. JP19H05614 and No. JP19H05615). The authors would like to thank S. Kawabata, K. Inomata, Y. Matsuzaki, and Y. Hashimoto for useful discussions, and R. Takano for the preliminary study of pulse shaping.

## Author contributions
N.T. conceived the idea, designed the circuits, performed the numerical simulations (except for pulse shaping) and experiments, and wrote the manuscript. T.Y. performed the numerical simulation of pulse shaping. T.Y., T.Y., and N.Y. supported theoretical aspects. All the authors discussed the results and reviewed the manuscript.

## Competing interests
The authors declare no competing interests.

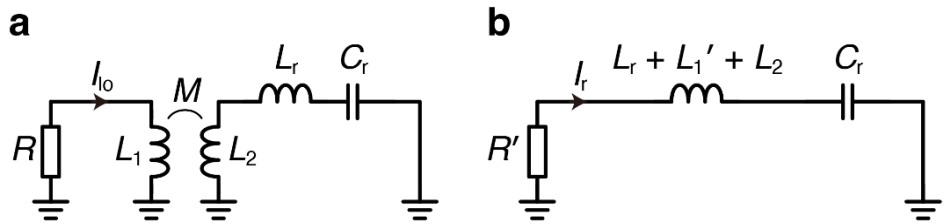

**Supplementary Fig. 1 | Resonator model. a.** Resonator coupled to the LO line with impedance $R$ through mutual inductance $M$. **b.** Equivalent circuit after converting the LO line using $M$ as an impedance inverter. From this figure, the resonance frequency and appropriate input power for each resonator in the AQFP-mux QC chip were estimated.